\documentclass[twocolumn,showpacs,preprintnumbers,amsmath,amssymb]{revtex4}
\usepackage{graphics}
\usepackage{epsfig}
\usepackage{bm}
\usepackage{umlaut}

\begin{document}
\title{Lattice dynamics and phonon softening in Ni-Mn-Al Heusler alloys}

\author{Xavier Moya, Llu\'{i}s Ma\~nosa}
\email{lluis@ecm.ub.es}
\author{Antoni Planes}

\affiliation{Departament d´Estructura i Constituents de la
Mat\`eria, Facultat de F\'isica, Universitat de Barcelona,
Diagonal 647, E-08028 Barcelona, Catalonia, Spain}

\author{Thorsten Krenke and Mehmet Acet}
\affiliation{Fachbereich Physik, Experimentalphysik, Universität
Duisburg-Essen, D-47048 Duisburg, Germany}

\author{V. O. Garlea, T. A. Lograsso, D. L. Schlagel and J. L. Zarestky}
\affiliation{Ames Laboratory, Department of Physics, Iowa State
University, Ames, Iowa 50011}
\date{\today}

\begin{abstract}
Inelastic and elastic neutron scattering have been used to study a
single crystal of the Ni$_{54}$Mn$_{23}$Al$_{23}$ Heusler alloy
over a broad temperature range. The paper reports the first
experimental determination of the low-lying phonon dispersion
curves for this alloy system. We find that the frequencies of the
TA$_2$ modes are relatively low. This branch exhibits an anomaly
(dip) at a wave number $\xi_{0} =\frac{1}{3}\approx 0.33$, which
softens with decreasing temperature. Associated with this
anomalous dip at $\xi_{0}$, an elastic central peak scattering is
also present. We have also observed satellites due to the magnetic
ordering.

\end{abstract}


\maketitle

\section{\label{sec:level1}Introduction}

\begin{figure*}
\includegraphics[width=0.9\linewidth,clip=]{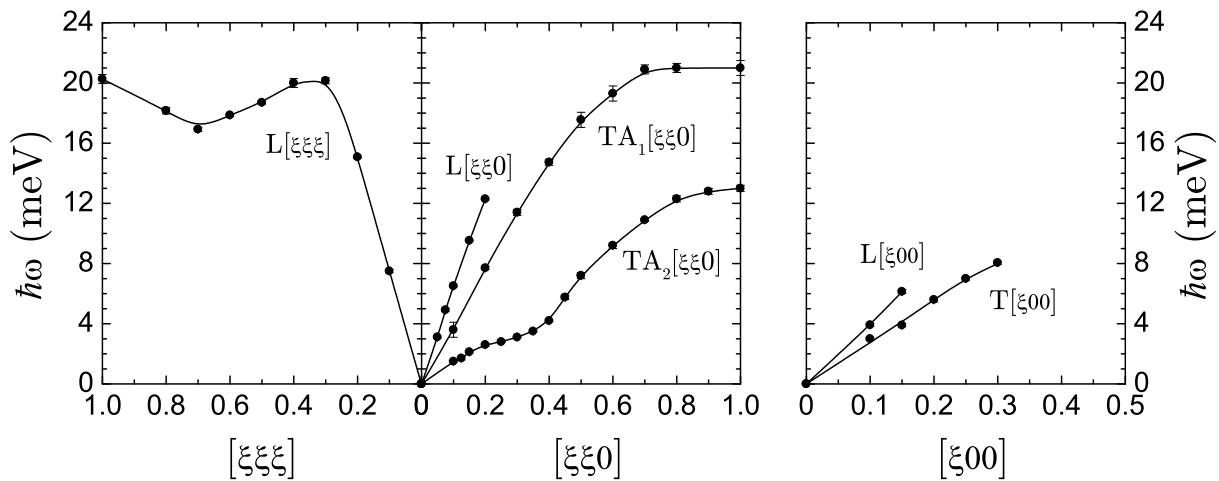}
\caption{Acoustic phonon dispersion curves along the high-symmetry
directions $[\xi\xi\xi]$, $[\xi\xi 0]$, and $[\xi 00]$, measured
at room temperature. Note the change of the horizontal scale
corresponding to the branch measured along the high-symmetry
direction $(\xi 00)$. Solid lines are guides to the eye.}
\label{fig1}
\end{figure*}

Magnetic shape memory behaviour is associated with the occurrence
of a structural (martensitic) transition in a magnetic material.
As a consequence of the degeneracy of the low temperature phase,
the martensitic state is a multidomain structure. Rotation of the
structural domains due to a difference in Zeeman energy occurs
under application of applied magnetic field \cite{OHandley98},
thus giving rise to a macroscopic deformation of the sample.
Strains as large as 10\% have been reported in single crystals of
Ni-Mn-Ga \cite{Ullakko1996,Sozinov2002}.

Neutron scattering experiments
\cite{Zheludev95,Zheludev96,Stuhr97,Manosa01} and elastic constant
measurements \cite{Worgull96,Manosa97,Stipcich04} have provided
evidence of an unusual lattice dynamical behaviour in the
prototypical magnetic shape memory alloy Ni-Mn-Ga. It has been
shown that the transverse TA$_2$ ($[\xi\xi0]$ direction,
$[\xi\overline{\xi}0]$ polarization) branch exhibits a dip
(anomalous phonon) at a specific wavenumber which depends on the
particular stacking sequence of the martensite structure. On
cooling the energies of the anomalous phonon decrease, and such a
phonon softening is enhanced when the sample orders
ferromagnetically. For compositions close to the stoichiometric
Ni$_2$MnGa, the energy of the anomalous phonon increases on
further cooling below a given temperature T$_I$, slightly higher
than the martensitic transition temperature T$_M$. Such a change
in behaviour is associated with a phase transition to the
so-called intermediate or pre-martensitic phase. This transition
is driven by a magnetoelastic coupling \cite{Planes97,Castan99}.
The temperature behaviour of the energy of the anomalous phonon
parallels that of the shear elastic constants, which also exhibit
a minimum at $T_I$ \cite{Manosa97,Stipcich04}.

The lattice dynamics of Ni-Mn-Ga alloys has been extensively
studied from first-principles calculations, and the main
experimental features have been successfully reproduced
\cite{Lee02,Zayak03,Bungaro03}. Recently, \textit{ab initio}
calculations have been extended to encompass other Heusler alloys
\cite{Enkovaara03,Busgen04,Zayak05}. Particular effort has been
devoted to computing the lattice dynamics of the Ni-Mn-Al system
over a broad composition range \cite{Busgen04}. For this system,
phonon dispersion curves along the $[110]$ direction have been
computed. A complete softening of the TA$_2$ phonons in the range
between $\xi = 0.25$ and $\xi = 0.4$ has been found. The shear
elastic constant $C'$ derived from the initial slope of the TA$_2$
branch has a lower value than the remaining constants.

Up to now, experimental data on phonon dispersion curves and
elastic constants in Heusler magnetic shape memory alloys are only
available for the Ni-Mn-Ga alloy system. The goal of the present
work is to present the first experimental determination of the
lattice dynamics of Ni-Mn-Al. Although Ni$_2$MnAl does not
transform martensitically, off-stoichiometric alloys transform
martensitically within a certain range of compositions close to
the stoichiometric one \cite{Kainuma96,Morito98}. The low
temperature martensitic structure depends on composition and the
observed structures are the same than those reported for other
Ni-Mn based magnetic shape memory alloys
\cite{Kainuma96,Morito98,Morito96,Pons00,Krenke05}. The magnetic
state of Ni$_2$MnAl consists of a mixed phase L2$_1$+B2, which
incorporates ferromagnetic and conical antiferromagnetic parts
\cite{Acet02,Manosa03,Ziebeck75}. This mixed state is due to the
significantly lower B2--L2$_1$ disorder-order transition
temperature, compared to that of Ni-Mn-Ga, which results in very
low kinetics for the ordering process \cite{Manosa03}. It is also
worth mentioning that large magnetic-field-induced strains have
been measured on this alloy system \cite{Fujita00}.

In this paper, we present elastic and inelastic neutron scattering
experiments in a Ni-Mn-Al single crystal. The measured phonon
dispersion curves and derived elastic constants are compared to
results obtained from \textit{ab initio} calculations
\cite{Busgen04}. Significant TA$_2$ phonon softening at $\xi_{0} =
0.33$ has been observed. In addition, a number of elastic
satellites associated with magnetic ordering and with structural
instabilities have been observed.

\section{Experimental Details}

The single crystal studied was grown by the Bridgman method.
Appropriate quantities of nickel (99.99 \% pure), aluminum (99.99
\%) and electrolytic manganese (99.9 \%) were cleaned and arc
melted several times under an argon atmosphere. The nominal
composition of the alloy was Ni$_{50}$Mn$_{25}$Al$_{25}$. The
button was then remelted and the alloy-drop cast into a copper
chill cast mold to ensure compositional homogeneity throughout the
ingot. The as-cast ingot was placed in alumina crucibles
(approximately 15 mm in diameter) and crystal growth was done in a
Bridgman-style refractory metal resistance furnace.  The ingot was
heated to 1350 ºC for 1 hr to allow homogenization before
withdrawing the sample from the heat zone at a rate of 5.0 mm/hr.
To minimize vaporization of the manganese during crystal growth,
the furnace was backfilled to a positive pressure of $6.8 \times
10^{5}$ Pa with purified argon after the chamber and sample had
been out gassed at 1350 ºC under vacuum. The average composition
of the alloy was determined by energy dispersive X-ray
photoluminescence analysis (EDX) to be Ni$_{54}$Mn$_{23}$Al$_{23}$
(within $\pm 1$ at. \%) .

Small pieces cut from the top and bottom of the ingot using a low
speed diamond saw were used as samples for magnetization and
calorimetric studies in order to characterize the sample. Both
magnetization and calorimetric studies revealed a magnetic
transition at $T \simeq 300$ K. Similar behaviour was observed in
polycrystalline samples with composition close to the studied
sample \cite{Manosa03}.

Neutron scattering measurements were carried out on the HB1A (Ames
Laboratory PRT) spectrometer at the High Flux Isotope Reactor
(HFIR) of the Oak Ridge National Laboratory. The monochromator and
analyzer used the (002) reflection of pyrolitic graphite (PG) and
highly oriented PG filters (HOPG) were used to minimize
higher-order contaminations of the beam. The HB1A spectrometer
operates at a fixed incident energy of 14.6 meV which requires
that scans with energy transfers over $\sim 9$ meV be performed
using neutron-energy gain. However, much of the low energy (low-q)
range of each branch was measured using the higher resolution
neutron-energy loss mode. Collimations of 48'-40'-40'-136' were
used and all scans were performed in the constant-Q mode while
counting against neutron monitor counts. Two sample settings were
used. The sample was aligned to have either the $(001)$ or the
$(1\overline{1}0)$ planes coincide with the scattering plane of
the spectrometer.

A standard closed-cycle helium refrigerator (displex) was used for
measurements below room temperature and up to 350 K. The sample
was mounted under helium atmosphere in an aluminum sample
container attached to the cold finger of the displex. The high
temperature measurements (623 K) were performed with the sample
mounted in a 10$^{-5}$ Torr vacuum furnace.

\section{Experimental results and discussion}

\subsection{Phonon dispersion}

\begin{figure}
\includegraphics[width=0.7\linewidth,clip=]{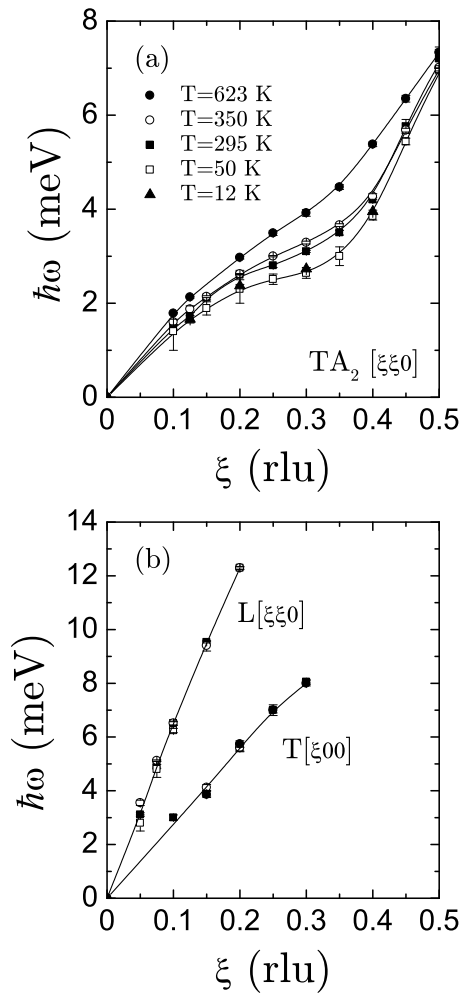}
\caption{Temperature dependence of the partial acoustic dispersion
curves. (a) Temperature dependence of the anomaly in the
TA$_2$[$\xi\xi 0$] branch. (b) Temperature dependence of the
longitudinal L[$\xi\xi0$] and transverse T[$\xi00$]. Whereas the
wiggle at  $\xi_{0}\approx 0.33$ in the TA$_2$[$\xi\xi 0$] branch
deepens with decreasing temperature, the other measured branches
do not change significantly. Solid lines are guides to the eye.}
\label{fig2}
\end{figure}

Phonon dispersion curves were determined from inelastic neutron
scattering along the high-symmetry directions $[\xi 00]$, $[\xi\xi
0]$, and $[\xi\xi\xi]$. The room temperature results are shown in
figure \ref{fig1}. Note that the dispersion curves are plotted in
an extended Brillouin zone scheme in the fcc L2$_1$ notation,
(lattice parameter $a= 5.82 \; \AA$ ). The phonon spectrum found
for the studied system shows the features typical of bcc materials
\cite{Planes01} : (i) low energies of the phonons in the
TA$_2$[$\xi\xi 0$] branch, and (ii) the marked dip at $\xi
=\frac{2}{3}$ in the longitudinal L[$\xi\xi\xi$] branch. The dip
in the L[$\xi\xi\xi$] branch is related to the incipient
instability towards the so-called $\omega$-phase. The most
noteworthy feature is the wiggle at $\xi_{0}\approx 0.33$ in the
TA$_2$[$\xi\xi 0$] branch, which deepens strongly with decreasing
temperature, as shown in figure \ref{fig2}(a). A similar behaviour
has also been reported for Ni-Mn-Ga alloys with composition close
to the stoichiometric Ni$_2$MnGa \cite{Zheludev95,Manosa01}.
According to recent \textit{ab initio} calculations
\cite{Busgen04}, such a phonon softening is due to a strong
electron-phonon coupling and the Kohn anomaly.

\begin{figure}
\includegraphics[width=0.9\linewidth,clip=]{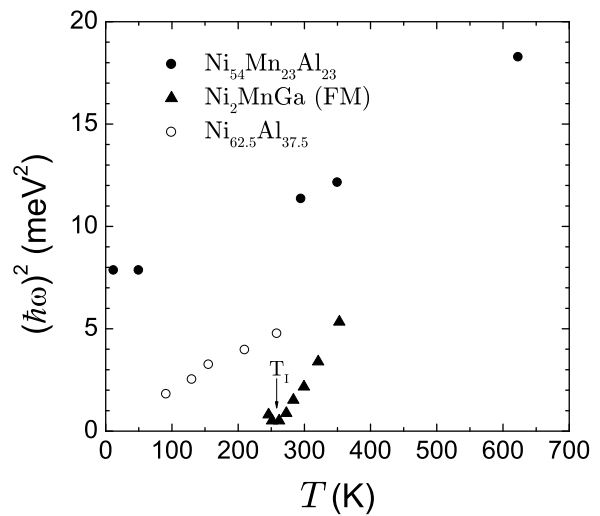}
\caption{Temperature dependence of the energy squared of the
TA$_2$[$\xi\xi 0$] $\xi_0=0.33$ phonon. Data from related systems
Ni$_2$MnGa and Ni$_{62.5}$Al$_{37.5}$ are also shown for
comparison. $T_I$ represents the premartensitic transformation
temperature. The data for the latter systems were taken
  from references \cite{Zheludev96} and \cite{Shapiro91} for Ni$_2$MnGa and Ni$_{62.5}$Al$_{37.5}$, respectively.}
\label{fig3}
\end{figure}

The temperature dependent softening of the  $\xi_{0} \simeq 0.33$
TA$_2$ phonon is illustrated in figure \ref{fig3}, which shows the
energy squared of the soft phonon as a function of temperature.
For comparison, data for the soft phonons in Ni$_2$MnGa and
Ni$_{62.5}$Al$_{37.5}$ alloys are also plotted. The degree of
softening is similar to that of Ni$_{62.5}$Al$_{37.5}$, although
the actual energy values are higher which is consistent with the
fact that the studied sample does not transform martensitically
within the studied temperature range. The softening in Ni$_2$MnGa
is stronger due to strong spin-phonon coupling in the
ferromagnetic state. It must be noted that the phonon branches
other than the TA$_2$ do not change significantly with temperature
as can be seen from figure \ref{fig2}(b).

The measured dispersion curves are in good agreement with those
obtained from \textit{ab initio} calculations for the [110]
direction of the stoichiometric Ni$_2$MnAl
\cite{Busgen04,Zayak05}. The energies of the acoustic phonons
coincide within the errors except for the low energy transverse
TA$_2$ branch, which exhibits complete softening  in the range
between $\xi =0.25$ and $\xi =0.4$. Experimental data are in
agreement with such an instability and show that the minimum is
located at $\xi_{0}\simeq 0.33$. However, from figure
\ref{fig2}(a), it can be seen that the softening is not complete,
i.e., the phonon frequency remains finite even  at the lowest
temperatures.

\begin{table}\caption{\label{tab:Elastic Constants} Elastic constants obtained from the
initial slopes of the acoustic phonon branches at $\xi \rightarrow
0$ at room temperature. The calculated values, corresponding to
the stoichiometric Ni$_2$MnAl sample, are extracted from reference
\cite{Busgen04}.}
\begin{ruledtabular}
\begin{tabular}{cccc}
& & \text{Measured value (GPa)} & \text{Calculated value (GPa)}\\
\hline
& $C_{44}$ & $103 \pm 5$ & 102\\
& $C_{L}$ & $259 \pm 5$ & 263\\
& $C^{'}$ & $16 \pm 5$ & 32\\
\end{tabular}
\end{ruledtabular}
\end{table}

The elastic constants at room temperature obtained from the
initial slopes of the acoustic phonon branches at $\xi \rightarrow
0$ are summarized in table \ref{tab:Elastic Constants}.
First-principles calculations of lattice dynamics have provided
theoretical values at 0 K for the elastic constants of the
stoichiometric Ni$_2$MnAl compound \cite{Busgen04}. The measured
and the calculated values are in good agreement, except for
$C^{'}$ which corresponds to the initial slope of the TA$_2$
branch. The value obtained from the measured curve is lower than
that obtained from the calculated one. While it is not expected
that $C_{L}$ and $C_{44}$ strongly depend on composition, $C^{'}$
may change from one composition to another. In addition, it is
worth noticing that the value is affected by a large error due to
the difficulty in defining the initial slope of the curve.

\subsection{Quasielastic scattering}

\begin{figure}
\includegraphics[width=0.7\linewidth,clip=]{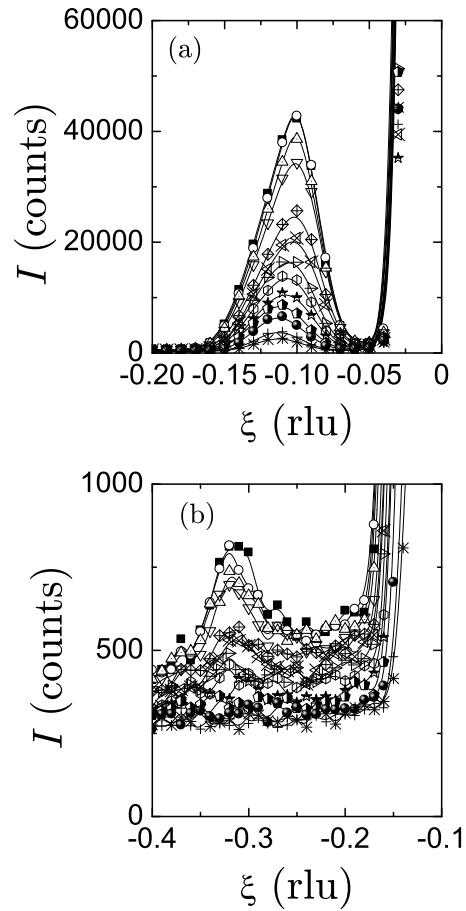}
\caption{Temperature dependence of the elastic scattering along
the [$\overline{\xi} \xi 0$] direction starting from the (020)
reflection. (a) Satellite due to the conical antiferromagnetic
order. (b) Central peak associated with the anomalous dip at
$\xi_0$ in the acoustic TA$_2$[$\xi\xi 0$] phonon branch.
Temperatures from top to bottom are 12 K, 50 K, 75 K, 100 K, 150
K, 175 K, 200 K, 225 K, 250 K, 275 K, 300 K, 330 K, and 350 K.
Solid lines are guides to the eye.} \label{fig4}
\end{figure}
\begin{figure}
\includegraphics[width=0.7\linewidth,clip=]{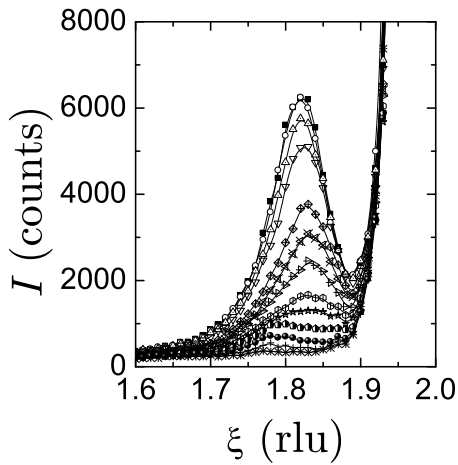}
\caption{Temperature dependence of the elastic scattering along
the [$\overline{\xi} \xi 0$] direction starting from the (220)
reflection. Temperatures from top to bottom are 12 K, 50 K, 75 K,
100 K, 150 K, 175 K, 200 K, 225 K, 250 K, 275 K, 300 K, 330 K, and
350 K. Solid lines are guides to the eye.} \label{fig5}
\end{figure}
\begin{figure}
\includegraphics[width=0.7\linewidth,clip=]{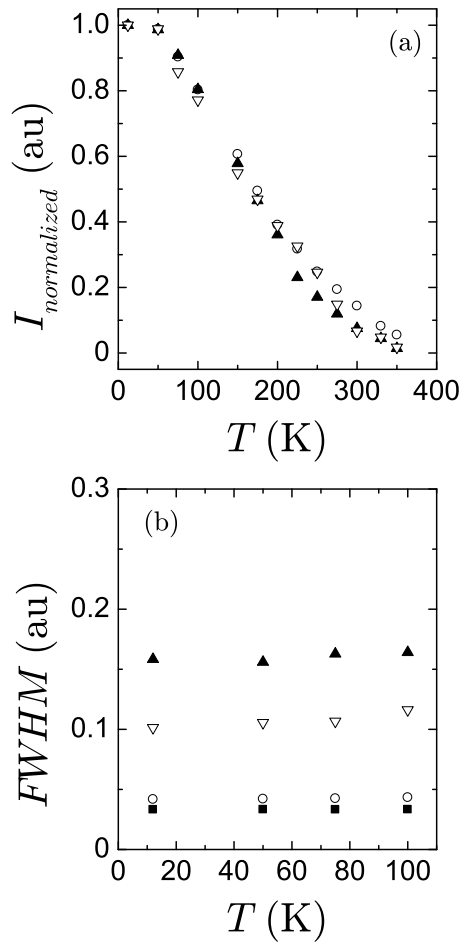}
\caption{Temperature dependence of (a) normalized maximum
intensity, and (b) width (full width at half maximum,
\textit{FWHM}) of the elastic satellites reported. Symbols
$\circ$, $\blacktriangle$, and $\triangledown$ represent the
$[-0.1,2.1,0]$, $[1.82,2.18,0]$, and $[-0.33,2.33,0]$ satellites,
respectively. The width of the (020) Bragg reflection
($\blacksquare$) is also shown in order to compare the widths of
the satellites to the fundamental reflections. None of the widths
of the Bragg peaks measured were instrumentally
limited.}\label{fig6}
\end{figure}

As previously reported in other related systems \cite{Zheludev95},
diffuse quasielastic scattering exists associated with the phonon
anomalies on the low-lying branch. A satellite peak -- the
\textit{central peak} -- develops at the wave vector which
corresponds to the position of the dip in the TA$_2$[$\xi\xi 0$]
branch. Figure \ref{fig4} shows the temperature dependence of the
elastic scattering along [$\overline{\xi} \xi  0$] direction
starting from the (020) reflection. From this figure, we can
distinguish two different elastic satellites in addition to the
Bragg peak at $\xi = 0$. The first satellite (fig. \ref{fig4}(a)),
at $\xi \approx 0.1$, is due to the conical antiferromagnetic
order as determined by Ziebeck and Webster \cite{Ziebeck75}. The
second satellite (fig. \ref{fig4}(b)), at $\xi \approx 0.33$, is a
central peak associated with the anomalous dip at $\xi_0$ in the
acoustic TA$_2$[$\xi\xi 0$] phonon branch.

We have also measured elastic scattering  along [$\overline{\xi}
\xi 0$] direction starting from the fundamental reflection (220).
Results are shown in Figure \ref{fig5}. A satellite located at
$\xi \approx 0.18$ is observed. Such a satellite was also reported
in Ni$_2$MnGa from X-ray \cite{Fritsch94} and neutron scattering
\cite{Stuhr97} experiments, and it was attributed to the splitting
of the Bragg peak owing to a tetragonal phase distortion.

Figure \ref{fig6}(a) shows the temperature dependence of the
maximum intensity of the observed satellites normalized to the
value at $T=12$ K. Figure \ref{fig6}(b) shows the temperature
dependence of the width at half maximum (\textit{FWHM}) of the
different satellites compared to that of the (020) Bragg
reflection. The satellite corresponding to the conical
antiferromagnetic order has similar width compared to the Bragg
reflections, whereas the other two reported satellites have less
intensity (see figs. \ref{fig4} and \ref{fig5}) and are broader in
q-space than the lattice reflections.

From the similarity in the temperature dependence of the $\xi
\approx 0.18$ and $\xi \approx 0.33$ satellites, Stuhr \textit{et
al.} \cite{Stuhr97} concluded that they were caused by dynamic
precursor effects rather than (static) contributions of the low
temperature phase. Our data for Ni-Mn-Al also show a similar
temperature dependence for these two satellites and therefore the
origin could be the same than in Ni-Mn-Ga: the $\xi \approx 0.18$
being a precursor of the tetragonal deformation and the $\xi
\approx 0.33$, a precursor of the martensite modulation. It must
be noted, however, that although the $\xi \approx 0.33$ satellite
has been observed in all the investigated Ni-Mn-Ga samples, the
one at $\xi \approx 0.18$ has not always been observed. It seems
that the appearance of such a satellite can be sample dependent
and it might be related to the homogeneity of the sample under
study. This is consistent with the fact that the martensitic
transition temperature in these alloy systems is extremely
sensitive on composition, and therefore the existence of a small
amount of martensite growing as temperature decreases cannot
completely disregarded.

Figures 4-6 show that precursor satellites only grow once the
sample orders magnetically. Again, this situation parallels the
behaviour observed in Ni-Mn-Ga. For that alloy system satellites
also grow below the Curie point \cite{Stuhr97}. Moreover, for that
alloy system, an enhancement of the phonon softening occurs in the
ferromagnetic phase \cite{Stuhr97,Manosa01}. Hence, present
results suggest that in Ni-Mn-Al the development of magnetic order
also affects lattice stability.

\section{Summary and conclusions}

We have measured the low-lying phonon dispersion curves of a
Ni$_{54}$Mn$_{23}$Al$_{23}$ alloy. We found that the frequencies
of the TA$_2$[$\xi\xi 0$] modes are relatively low thus evidencing
a low dynamical stability for distortions on the $\{110\}$ planes
along the $\langle 1 \bar{1} 0 \rangle$ directions. This branch
has an anomaly (dip) at a wave vector $\xi_{0} =\frac{1}{3}\approx
0.33$. The energy of this phonon decreases with decreasing
temperature. The existence of this anomaly in the phonon branch
was predicted from recent \textit{ab initio} calculations.
Associated with this anomalous dip at $\xi_{0}$, an elastic
central peak scattering is also present. The reported behaviour is
similar to that observed in the related Ni-Mn-Ga system but with a
less softening of the TA$_2$ phonon in Ni-Mn-Al. Overall, the
energies of the acoustic TA$_2$ branch show higher values compared
to those in Ni-Mn-Ga. This is consistent with the fact that the
studied crystal does not transform martensitically within the
studied temperature range.

\begin{acknowledgments}

This work received financial support from the CICyT (Spain),
Project No. MAT2004--01291, Marie--Curie RTN MULTIMAT (EU),
Contract No. MRTN--CT--2004--505226, DURSI (Catalonia), Project
No. 2005SGR00969, and from the Deutsche Forschungsgemeinschaft
(GK277). XM acknowledges support from DGICyT (Spain). Ames
Laboratory is operated for the U.S. Department of Energy by Iowa
State University under Contract No. W-7405- Eng-82. The work at
Ames was supported by the Director for Energy Research, Office of
Basic Energy Sciences.

\end{acknowledgments}


\bibliography{apssamp}

\end{document}